\documentclass{article}[9pt]

\usepackage{spconf,amsmath,graphicx,epsfig,multicol,url}
\usepackage[utf8]{inputenc}
\usepackage{multirow}
\usepackage{booktabs}
\usepackage{dcolumn}
\usepackage{enumitem}
\usepackage{tikz}
\usetikzlibrary{shapes,arrows}
\usepackage{color}

\newcommand{\EE}       {\boldsymbol{\Sigma}}
\newcommand{\tmatrix}  {\mathbf{T}}
\newcommand{\mm}       {\boldsymbol{\mu}}

\newcommand{\bphi}{\boldsymbol{\phi}}

\title{Discriminatively Re-trained i-vector Extractor for Speaker Recognition}
\name{ Ond\v{r}ej Novotn\'y, Old\v{r}ich Plchot, Ond\v{r}ej Glembek, Luk\'a\v{s} Burget, Pavel Mat\v{e}jka\thanks{The work was supported by Czech Ministry of Interior project No. VI20152020025 "DRAPAK", Google Faculty Research Award program, Czech Science Foundation under project No. GJ17-23870Y, and by Czech Ministry of Education, Youth and Sports from the National Programme of Sustainability (NPU II) project "IT4Innovations excellence in science - LQ1602".}}
\address{
Brno University of Technology, Speech@FIT and IT4I Center of
Excellence, Brno, Czechia \\
\texttt{inovoton@fit.vutbr.cz}
}
\begin{document}
\ninept
\maketitle
\begin{abstract}
%In this work we present a possible promising practical pervert peculiar approach to discriminative training 
In this work we revisit discriminative training of the i-vector extractor component in the standard speaker verification (SV) system.  The motivation of our research lies in the robustness and stability of this large generative model, which we want to preserve, and focus its power towards any intended SV task. We show that after generative initialization of the i-vector extractor, we can further refine it with discriminative training and obtain i-vectors that lead to better performance on various benchmarks representing different acoustic domains. 
\end{abstract}
\section{Introduction}
\label{sec:intro}

In recent years, there have been many attempts to take advantage of neural
networks (NNs) in speaker verification. 
Most of the attempts have replaced or improved one of the components
of an i-vector + PLDA system (feature extraction, calculation of sufficient statistics, i-vector extraction or PLDA) with a neural network. As examples, let us mention: using NN bottleneck features instead of conventional MFCC features \cite{lozano_odyssey_2016}, NN acoustic models replacing Gaussian Mixture Models for extraction of sufficient statistics~\cite{Lei_icassp_2014}, NNs for either complementing PLDA \cite{Novoselov_interspeech_2015,Bhattacharya_SLT16} or replacing it \cite{Ghahabi_icassp_2014}. More ambitiously, NNs that take the frame level features of an utterance as input and directly produce an utterance level representation---usually referred to as an \emph{embedding}---have in the past two years almost replaced the generative i-vector approach in text independent speaker recognition~\cite{Variani_icassp_2014, heighold_icassp_2016, zhang_slt_2016, snyder_slt_2016, Bhattacharaya_interspeech_2017, snyder_interspeech_2017, xvec:Snyder2018}. 

These embeddings are obtained by the means of \emph{pooling mechanism}, for example taking the mean, over the frame-wise outputs of one or more layers in the NN~\cite{Variani_icassp_2014}, or by the use of a recurrent NN~\cite{heighold_icassp_2016}.  An obvious advantage---compared to i-vectors---lies in a much smaller amount of model parameters, which is typically around 10 million in the \emph{x-vector} case~\cite{snyder_interspeech_2017, xvec:Snyder2018} compared to the i-vector with approximately 50 million parameters for both UBM and i-vector extractor. This results in a very fast and memory efficient embedding extraction. A disadvantage of the x-vector framework can be seen in training during which it is essential to massively augment the training data and split them into many rather short (2--5 seconds) examples.
%, which effectively acts as a kind of regularization. 

In this research, we keep the large parameter space from the generative i-vector extractor and we focus on discriminative retraining of such a model that still uses a fairly complex GMM-UBM to provide the training examples (sufficient statistics). I-vector model is generally very robust, which is a property that we want to retain, but at the same time we want the model to focus on important features with respect to the task at hand---discrimination between speakers---and at the same time do not waste parameters to represent the redundant variability in the data.

To obtain a standalone discriminative i-vector extractor, we used the same strategy as in the x-vector framework~\cite{Variani_icassp_2014, Bhattacharaya_interspeech_2017, snyder_interspeech_2017} and we retrained the NN representation of our generative model to optimize the multi-class cross-entropy over a set of training speakers.  
This is in contrast with our previous research~\cite{dix:glembek}, where we optimized the binary cross-entropy over verification trials formed by pairs of i-vectors. We show that with such an approach we can achieve a reasonable improvement in performance.  Our results are perhaps not as good as what can be achieved with current x-vector systems~\cite{xvec:ondran}, but our goal is to further use this model in the fully end-to-end discriminative system~\cite{rohdin:icassp:2018} that can be initialized from a robust generative baseline.  In~\cite{rohdin:icassp:2018}, we were already able to build such a system, but it was just the i-vector extractor component that posed the biggest challenge and we had to resort to ad-hoc simplifications like PCA-based dimensionality reduction of input sufficient statistics. % from the GMM-UBM. 

In order to compare both approaches (generative and discriminative) on speaker verification task, both versions of i-vectors were extracted and used in a standard generative PLDA backend.

%Such systems have recently been proven competitive for both short and long utterance durations in text-independent speaker verification \cite{Bhattacharaya_interspeech_2017, snyder_interspeech_2017}.

%\begin{itemize}[nosep]
%   % \item \cite{rohdin:icassp:2018}, napsat, ze lze to pouzit primo v end-to-end misto snahy to tam aproximovat
%    \item \cite{dix:glembek}, \textcolor{red}{@@ odkaz na ondrovu predchozi praci nekde zakomponovat}
%    %\item smerovat to spis asi jako proof of concept
%    As opposed to~\cite{dix:glembek}, where the discriminative training was based 
%    on optimizing a 2-class cross-entropy of recognizing pairs of i-vectors, in this work we ...
%\end{itemize}

\section{Theoretical Background}
\label{sec:SRE}

The i-vectors~\cite{DehakN_TASLP:2010} provide a way of reducing
large-dimensional input data to a small-dimensional feature 
vector while retaining most of the relevant information. 
The main principle is that the utterance-dependent Gaussian 
Mixture Model (GMM) supervector of concatenated
mean vectors lies in a low-dimensional subspace---defined by matrix $\tmatrix$, commonly referred to as an \emph{i-vector extractor}---and whose coordinates are given by the i-vector $\bphi$. 
The closed-form solution for computing the i-vector can be expressed as a function of
the \emph{zero-} and \emph{first-order GMM statistics}:
$\mathbf{n}_\mathcal{X}=[N_{\mathcal{X}}^{(1)}, \ldots,N_{\mathcal{X}}^{(C)}]'$
and $\mathbf{f}_\mathcal{X}=[\mathbf{f}_{\mathcal{X}}^{(1)'},
\ldots,\mathbf{f}_{\mathcal{X}}^{(C)'}]'$, where
\begin{eqnarray}
\label{Eq:ZeroOrderStats}
N_{\mathcal{X}}^{(c)} & = & \sum_{t} \gamma_{t}^{(c)} \\[-2mm]
\label{Eq:FirstOrderStats}
\mathbf{f}_{\mathcal{X}}^{(c)} & = & \sum_{t} \gamma_{t}^{(c)} \mathbf{o}_{t} ,
\end{eqnarray}
where $\gamma_{t}^{(c)}$ is the posterior (or occupation) probability of frame
$t$ being generated by the mixture component $c$.  
% The tuple $\gamma_{t}=(\gamma_{t}^{(1)},\ldots,\gamma_{t}^{(C)})$ is usually referred to as \emph{frame alignment}.  
%
The i-vector is then computed as
\begin{equation}
\label{Eq:westimation}
\bphi_{\mathcal{X}} = \mathbf{L}_{\mathcal{X}}^{-1} \bar{\mathbf{T}}' \bar{\mathbf{f}}_{\mathcal{X}}
\end{equation}
with
%
% \vspace{-1mm}
\begin{equation}
\label{Eq:Lestimation}
\mathbf{L}_{\mathcal{X}} = \mathbf{I} + \sum_{c=1}^{C} N_{\mathcal{X}}^{(c)}
\bar{\mathbf{T}}^{(c)'}  \bar{\mathbf{T}}^{(c)} ,
\end{equation}
where $\bar{\mathbf{f}}_{\mathcal{X}}^{(c)}$ and $\bar{\mathbf{T}}^{(c)}$
are the ``normalized'' variants of $\mathbf{f}_{\mathcal{X}}^{(c)}$ and $\mathbf{T}^{(c)}$, respectively:
\begin{eqnarray}
\label{Eq:NormF}
\bar{\mathbf{f}}_{\mathcal{X}}^{(c)}  & = & \EE^{(c)-\frac{1}{2}} \left(\mathbf{f}_{\mathcal{X}}^{(c)} -  N_{\mathcal{X}}^{(c)} \mm^{(c)} \right ) \\[-1mm]
\label{Eq:NormT}
\bar{\mathbf{T}}^{(c)}  & = & \EE^{(c)-\frac{1}{2}} \mathbf{T}^{(c)},
\end{eqnarray}
and $\EE^{(c)-\frac{1}{2}}$ is a symmetrical decomposition (such as Cholesky) 
of an inverse of the GMM UBM covariance matrix $\EE^{(c)}$.

%%  For convenience, we center the first-order statistics around the UBM means, which allows us to treat the UBM means effectively as a vector of zeros.
%%  
%%  \begin{eqnarray} 
%%  \label{Eq:NormStats}
%%  \mathbf{f}_{\mathcal{X}}^{(c)}  &\leftarrow&  \mathbf{f}_{\mathcal{X}}^{(c)} - N_{\mathcal{X}}^{(c)}\mathbf{m}^{(c)} \\
%%  \mathbf{m}^{(C)} &\leftarrow&  \mathbf{0}
%%  \end{eqnarray}
%%  
%%  Similarly, we ``normalize'' the first-order statistics and the matrix $\mathbf{T}$ by the UBM covariances, which again allows us to treat the UBM covariances as an identity matrix:
%%  %
%%  
%%  \begin{eqnarray} 
%%  \label{Eq:NormStats}
%%  \mathbf{f}_{\mathcal{X}}^{(c)} & \leftarrow &  \mathbf{\Sigma}^{(c)-\frac{1}{2}}       \mathbf{f}_{\mathcal{X}}^{(c)} \\
%%  \mathbf{T}^{(c)}  &\leftarrow&   \mathbf{\Sigma}^{(C)-\frac{1}{2}} \mathbf{T}^{(c)} \\
%%  \mathbf{\Sigma}^{(c)} &\leftarrow&  \mathbf{0},
%%  \end{eqnarray}
%%  
%%  where $\mathbf{\Sigma}^{(C)-\frac{1}{2}}$ is Cholesky decomposition of an inverse $\mathbf{\Sigma}$, and $\mathbf{T}^{(C)}$ is an $F \times M$ submatrix of $\mathbf{T}$ corresponding to $c$ mixture component such that $ \mathbf{T} = \left ( \mathbf{T}^{(1)} ,\dots, \mathbf{T}^{(C)} \right)$
%%  
%%  % \subsection{i-vector extraction}

%%%%%%%%%%%%%%%%%%%%%%%%%%%%%%%%%%%%%%%%%%%%%%%%%%%%%%%%%%%%%%%%%%%%%%%%%%%%%%%%
%%%%%%%%%%%%%%%%%%%%%%%%%%%%%%%%%%%%%%%%%%%%%%%%%%%%%%%%%%%%%%%%%%%%%%%%%%%%%%%%

\subsection{Discriminatively Trained i-vector Extractor}
\label{sec:DiX}

Traditionally, matrix $\tmatrix$ is trained in a 
generative fashion using the EM algorithm.  In this work, however, we focus on the i-vector extractor parameter re-estimation to better discriminate between speakers.  Our experimental pipeline is plotted in Fig.~\ref{sec:intro}.

Multi-class logistic regression was used as a classifier, where the posterior probability of class (speaker) $k$ given i-vector $\bphi_{\mathcal{X}_n}$ (as defined in Eq.~(\ref{Eq:westimation})) is computed as:
\begin{equation} 
\label{Eq:cross}
p_{\mathbf{W}}(C_{k} \mid \bphi_{\mathcal{X}_n})  =  \frac{\exp( \mathbf{w}_{k}^T\bphi_{\mathcal{X}_n})}{\sum_{j} \exp( \mathbf{w}_{j}^T\bphi_{\mathcal{X}_n})}, 
\end{equation}
where $\mathbf{W}=[\mathbf{w}_{1},\ldots,\mathbf{w}_{K}]$ are the parameters of
logistic regression.  
% Input of the network  are normalized statistic $\bar{\mathbf{f}}_{\mathcal{X}}^{(c)}$ 
% and  $N_{\mathcal{X}}^{(c)}$

Multi-class cross-entropy was used as the objective function:
\begin{eqnarray} 
\label{Eq:crossE}
%E(\mathbf{w_1},\dots,\mathbf{w_K}, \mathbf{T}) = -\sum^{N}_{n=1} \sum^{K}_{k=1} s_{nk} p(C_{k} \mid \bphi_{\mathcal{X}_n}),
E(\mathbf{W}, \mathbf{T}) = -\sum^{N}_{n=1} \sum^{K}_{k=1} s_{nk} p_{\mathbf{W}}(C_{k} \mid \bphi_{\mathcal{X}_n}),
\end{eqnarray}
%
%%
%\begin{eqnarray} 
%\label{Eq:cross}
%%E(\mathbf{w_1},\dots,\mathbf{w_K}, \mathbf{T}) = -\sum^{N}_{n=1} \sum^{K}_{k=1} s_{nk} p(C_{k} \mid \bphi_{\mathcal{X}_n}),
%E_{LR}(\mathbf{W}, \mathbf{T}) = - \sum^{K}_{k=1} s_{nk} p_{\mathbf{W}}(C_{k} \mid \bphi_{\mathcal{X}_n}),
%\end{eqnarray}
%
where, $s_{nk}$ is $k$-th 
element of the target variable in 1-of-K coding, $K$ is number of speakers 
(classes), and $N$ is number of training samples.  For the purpose of this work, 
let us treat
the i-vector $\bphi_{\mathcal{X}_n}$ as a function of $\tmatrix$.
In stage-1, we trained the classifier (parameters $\mathbf{W}$) only.  
After several epochs (stage-2), we trained the classifier and $\tmatrix$-matrix jointly. 
As the optimizer, stochastic gradient descent algorithm was used.
%After the i-vector extractor retraining, the standard pipeline for extraction is used.

% Let us note that multi-class logistic-regression can also be understood as a relatively simple means of retraining and boosting the i-vector extractor module and preparing it for the final task---the end-to-end system---or replace logistic-regression by discriminative PLDA.

\begin{figure}[tb]
  \centering
% \begin{tikzpicture}[
% roundnode/.style={circle, draw=green!60, fill=green!5, very thick, minimum size=7mm},
% squarednode/.style={rectangle, draw=red!60, fill=red!5, very thick, minimum size=5mm}
% 
% 
% ]
% 
%  
% \draw
% 	% Drawing the blocks of first filter :
% 	node at (0,0)[right=-3mm]{xxx}
% 	node [input, name=input1] {} 
% 	node [sum, right of=input1] (suma1) {cosi}
% 	node [block, right of=suma1] (inte1) {cosi}
%          node at (6.8,0)[block] (Q1) {\Large $Q_1$}
%          node [block, below of=inte1] (ret1) {\Large$T_1$};
%     % Joining blocks. 
%     % Commands \draw with options like [->] must be written individually
% 	\draw[->](input1) -- node {$X(Z)$}(suma1);
%  	\draw[->](suma1) -- node {} (inte1);
% 	\draw[->](inte1) -- node {} (Q1);
% 	\draw[->](ret1) -| node[near end]{} (suma1);
% 	% Adder
% \end{tikzpicture}
  \scalebox{1.0}{
  \includegraphics[width=1.0\linewidth]{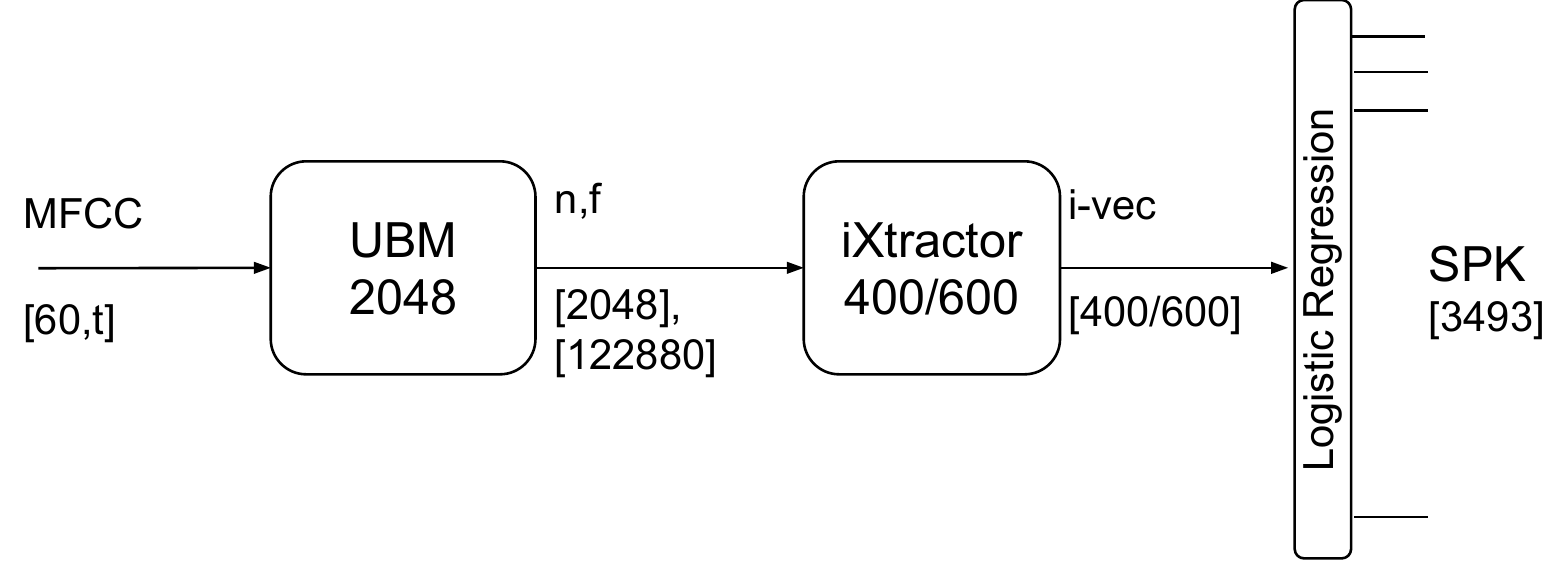}}
  \caption{Training pipeline of i-vector extractor parameters re-estimation. During the initial phase of training, only the logistic regression is trained. During the second phase, the parameters of the logistic regression and the i-vector extractor (iXtractor) are updated.}
  \label{fig:iXpipeline}
  % \vspace{-5mm}
\end{figure}

%%%%%%%%%%%%%%%%%%%%%%%%%%%%%%%%%%%%%%%%%%%%%%%%%%%%%%%%%%%%%%%%%%%%%%%%%%%%%%%%
%%%%%%%%%%%%%%%%%%%%%%%%%%%%%%%%%%%%%%%%%%%%%%%%%%%%%%%%%%%%%%%%%%%%%%%%%%%%%%%%

\section{System Setup}

%%%%%%%%%%%%%%%%%%%%%%%%%%%%%%%%%%%%%%%%%%%%%%%%%%%%%%%%%%%%%%%%%%%%%%%%%%%%%%%%
%%%%%%%%%%%%%%%%%%%%%%%%%%%%%%%%%%%%%%%%%%%%%%%%%%%%%%%%%%%%%%%%%%%%%%%%%%%%%%%%

\subsection{Datasets}

We used the PRISM~\cite{ferrer:sre11} training dataset definition without added 
noise or reverberation to train UBM and i-vector extractor. 
The  set comprises Fisher 1 and 2, Switchboard phase 2 and 3 and Switchboard cellphone phases 
1 and 2, along with a set of Mixer speakers. This includes the 66 held out 
speakers from SRE10 (see Section III-B5 of~\cite{ferrer:sre11}), and 965, 980, 485 and 310 speakers 
from SRE08, SRE06, SRE05 and SRE04, respectively. A total of 13,916 speakers 
are available in Fisher data and 1,991 in Switchboard data.

Two variants of gender-independent PLDA models were trained: one on the clean training data, the second 
included also artificially added different mixes of noises and reverberation. 
Artificially added noise and reverb segments totaled approximately $24000$ segments or $30\%$ of total number of clean segments for PLDA training, see details in Sec.~\ref{lab:pldaaugset}.

We evaluated our systems on the \emph{female} portions
of NIST SRE 2010~\cite{NIST_SRE:WWW} ({\it tel-tel}, {\it int-int} and {\it int-mic}) and PRISM ({\it prism,noi}, {\it prism,rev} and {\it prism,chn}, see section III.B of~\cite{ferrer:sre11}), where {\it tel-tel} and {\it prism,chn} represent telephone speech, {\it int-int} and {\it int-mic} interview speech and {\it prism,noi} with {\it prism,rev} represent artificially corrupted speech with noise and reverberation.

%We evaluated our systems on the \emph{female} portions of the following conditions 
%of NIST SRE 2010~\cite{NIST_SRE:WWW} and PRISM:
%
%\begin{itemize}
% \item {\it tel-tel}: SRE 2010 extended telephone condition 
%involving normal vocal effort conversational telephone speech in enrollment and 
%test (known as ``condition 5'').
%  \item {\it int-int}: SRE 2010 extended interview condition involving 
%interview speech from different microphones in enrollment and test (known as 
%``condition 2'').
%  \item {\it int-mic}: SRE 2010 extended interview-microphone condition 
%involving interview enrollment speech and normal vocal effort conversational 
%telephone test speech recorded over a room microphone channel (known as 
%``condition 4'').
% \item {\it prism,noi}: Clean and artificially noised waveforms from both interview and telephone conversations recorded over lavalier microphones.
% Noise was added  at different SNR levels and recordings tested against each other.
% \item {\it prism,rev}: Clean and artificially reverberated waveforms from both interview and telephone conversations recorded over lavalier microphones.
%Reverberation was added with different RTs and recordings are tested against each other.
%\item {\it prism,chn}: English telephone conversation with normal vocal effort 
%recorded over different microphones from both SRE2008 and 2010 are tested against each other.
%\end{itemize}
%
Additionally, we used the {\it Core-Core} condition from the SITW challenge---\emph{sitw-core-core}. SITW~\cite{SITW_evaluation_plan} dataset is a large collection of real-world data exhibiting speech from individuals across a wide array of challenging acoustic and environmental conditions. 

%These audio recordings do not contain any artificially added noise, reverberation or other artifacts. This database was collected from open-source media.  The \emph{sitw-core-core} condition comprises audio files each containing a continuous speech segment from a single speaker. Enrollment and test segments contain between 6-180 seconds of speech. We scored all trials (both genders). 

We also test on NIST SRE 2016~\cite{NIST:SRE2016}, but we split the trial set by language into Tagalog ({\it sre16-tgl-f})  and Cantonese ({\it sre16-yue-f}).  We use only female trials (both single- and multi-session). We did not use SRE'16 unlabeled development set in any way.

\subsection{PLDA and i-vector Extractor Augmentation Sets}
\label{lab:pldaaugset}
To extend the training set, we created new artificially corrupted training sets from the PRISM training set. In addition to using noise and reverberation presented below, data were also augmented with randomly generated cuts. In our experiments, we used 30\% of original training data to generate cuts with durations between 3 to 5 seconds.

% \vspace{-3mm}
\subsubsection{Adding Noise}
We prepared a dataset of noises from three different sources:
\begin{itemize} % [nosep]
\item 200 samples (4 minutes long) taken from the Freesound library\footnote{\url{http://www.freesound.org}} (real fan, HVAC, street, city, shop, crowd, library, office and workshop).
\item 5 samples (4 minutes long) of artificially generated noises: various spectral modifications of white noise + 50 and 100 Hz hum.
\item 18 samples (4 minutes long) of babbling noises by merging speech from 100 random speakers from Fisher database using speech activity detector. 
\end{itemize}

%\noindent
%Noises were divided into three disjoint groups for training (223 files), development (40 files) and test (41 files). Development and test subsets are not used in this work. 

% \vspace{-3mm}
\subsubsection{Reverberation}
The prepared set consists of real room impulse responses from several databases: MARDY~\cite{MARDY:WWW}, AIR~\cite{AIR:WWW}, C4DM~\cite{C4DM:WWW,C4DM:WWW2},  OPENAIR~\cite{OpenAir:WWW}, RVB 2014~\cite{RVB:WWW}, and RWCP~\cite{RWCP:WWW}. 
Together, they form a set with all types of rooms (small rooms, big rooms, lecture room, restrooms, halls, stairs etc.). All room models have more than one impulse response per room (different RIR was used for source of the signal and source of the noise to simulate different locations of their sources). Rooms were split into two disjoint sets, with 396 rooms for training and 40 rooms for test.

% \vspace{-3mm}
\subsubsection{Composition of the Augmented Training Set}
\label{section:signal_corruption}
To mix the reverberation, noise and signal at given SNR, we followed the  procedure outlined in \cite{xvec:ondran}.

When jointly augmenting the data by noise and reverberation, the speech and noise are reverberated separately and different RIRs from the same room are used for speech signal and noise to simulate different positions of their sources.
In the following step, we set a ratio of noise and signal energies to obtain the required SNR. Energies of the signal and noise are computed from frames given by the original signal's voice activity detection (VAD). Finally, signal and noise are summed together at desired SNR.
In case we want to add only noise or reverberation, the appropriate part of the algorithm is used.

\begin{table*}[!htb]
\centering
\caption{\label{tab:mainresults} Comparison of the i-vector baseline with different approaches used for i-vector extractor training.  Both blocks are divided into columns corresponding to the dimensionality of i-vectors (we used 400- and 600- dimensions). 
Results are also divided based on the training set of PLDA, where we used clean and multi-condition fashion (with noised and reverberated data.
Results (EER $[\%]$) in each column correspond to the different i-vector extractor training setup: B - generative baseline without augmented data, G- generative training with augmented data and D - augmented data used for discriminative retraining.}\vspace{2mm}
%\hskip-4.0cm
\scalebox{0.8}{
\begin{tabular}{ l  r r r  c  r r r  c  r r r  c  r r r }
\toprule
%\cline{2-13}
\addlinespace[0.05cm]
\multicolumn{1}{c}{} & \multicolumn{7}{c}{400-dim} & \multicolumn{1}{c}{} & \multicolumn{7}{c}{600-dim} \\ 
\addlinespace[0.05cm] 

\cmidrule(rl){2-8} \cmidrule(rl){10-16} 

\multicolumn{1}{c}{}  & 
\multicolumn{3}{c}{PLDA clean} & & \multicolumn{3}{c}{PLDA extension data} & & 
\multicolumn{3}{c}{PLDA clean} & & \multicolumn{3}{c}{PLDA extension data} \\ 

\cmidrule(rl){2-4} \cmidrule(rl){6-8} \cmidrule(rl){10-12} \cmidrule(rl){14-16} 

Condition & 
  \multicolumn{1}{c}{B} & \multicolumn{1}{c}{G} & \multicolumn{1}{c}{D} & &
  \multicolumn{1}{c}{B} & \multicolumn{1}{c}{G} & \multicolumn{1}{c}{D} & & 
  \multicolumn{1}{c}{B} & \multicolumn{1}{c}{G} & \multicolumn{1}{c}{D} & &
  \multicolumn{1}{c}{B} & \multicolumn{1}{c}{G} & \multicolumn{1}{c}{D} \\ 

\midrule
 tel-tel                &  2.23 &  2.43 &  \textbf{1.97} & &  3.36 &  3.73 &  3.25 & &  1.99 &  1.98 &  1.84 & &  2.74 &  2.86 &  2.70 \\
 sre16-yue-f            & 10.90 & 11.25 & 10.97 & & 11.32 & 11.20 & 10.87 & & 11.20 & 11.32 & 11.10 & & 11.53 & 11.20 & 11.17 \\
\midrule
 int-int                &  4.72 &  4.75 &  4.37 & &  4.83 &  4.88 &  4.56 & &  4.57 &  4.52 &  4.47 & &  4.55 &  4.71 &  4.52 \\
 int-mic                &  2.15 &  2.24 &  2.11 & &  2.02 &  2.28 &  1.91 & &  1.85 &  2.11 &  1.91 & &  2.00 &  2.02 &  1.94 \\
 prism,chn              &  1.13 &  1.48 &  \textbf{0.88} & &  1.14 &  1.40 &  1.14 & &  1.03 &  0.92 &  0.95 & &  0.97 &  1.04 &  0.94 \\
 sitw-core-core         & 10.51 & 10.75 & 10.29 & & 10.57 & 10.62 & 10.21 & & 10.11 & 10.28 &  9.82 & & 10.32 & 10.34 & 10.17 \\
\midrule
 prism,noi              &  4.43 &  4.51 &  3.97 & &  3.66 &  3.90 &  3.44 & &  3.72 &  3.79 &  3.58 & &  3.42 &  3.26 &  3.25 \\
 prism,rev              &  2.81 &  3.06 &  2.54 & &  2.45 &  2.47 &  2.34 & &  2.51 &  2.74 &  2.35 & &  2.23 &  2.22 &  2.15 \\

\bottomrule
\end{tabular}}
\end{table*}

%%%%%%%%%%%%%%%%%%%%%%%%%%%%%%%%%%%%%%%%%%%%%%%%%%%%%%%%%%%%%%%%%%%%%%%%%%%%%%%%
%%%%%%%%%%%%%%%%%%%%%%%%%%%%%%%%%%%%%%%%%%%%%%%%%%%%%%%%%%%%%%%%%%%%%%%%%%%%%%%%

\section{Experiments and Discussion}
\label{sec:experiments}

We conducted a set of experiments, all with three different approaches in i-vector extractor training, denoted with letters B, G, D further on. 
\begin{description} % [noitemsep]
\item [B] --- In the first set of experiments, we trained a baseline i-vector extractor in the traditional generative way, using the original PRISM training corpus. 
\item [G] --- In the second variant, we still used generative training, but we 
augmented the training data with noise, reverberation, and cuts as described in 
Section~\ref{lab:pldaaugset}.  
%In both variants, we used 10 iterations of EM algorithm.
%
\item[D] --- In the last variant, we used the pre-trained generative i-vector 
extractor from G and we retrained it discriminatively.  The training pipeline is 
shown in Fig.~\ref{fig:iXpipeline}.
\end{description}

For discriminative training, the data preparation was necessary to avoid classifier overtraining.  We used speakers with at least 5 utterances in original data only.  This step limits the training data to 3493 speakers with 59112 utterances (177336 utterances including augmentation). 

Our results (EER) are presented in Tab.~\ref{tab:mainresults}. 
The results are presented for 400- and 600-  dimensional i-vectors. Next, results are also divided based on PLDA, where we distinguish PLDA trained on clean data and multi-condition training. 
Finally, the table is also divided based on the type of the condition, for the telephone channel, microphone and artificially created conditions. We did not use any type of adaptation or any other technique used for results improvement in conditions from SRE16 and others. 

When we compare baseline systems (B column in the table) with discriminatively 
retrained variant (D column in the table), we can see that except two cases 
({\it sre16-yue-f} in 400-dim variant with ``clean'' PLDA and {\it int-mic} in  600-dim 
variant with ``clean'' PLDA) the discriminative training is better.  The 
discriminative approach is also better compared with the generative approach 
with augmented data in the training, where you can see that augmentation in 
generative training caused mostly degradation in the final. 

In {\it tel-tel} condition, we can see significant improvement with discriminative training, where 400-dim system have almost 12\% relative improvement compared to baseline and it also outperforms 600-dim baseline. The similar situation is in the {\it prism,ch} condition, where we have 22\% relative improvement in 400-dim variant, here we also outperform 600-dim discriminative variant of training.

The general improvement can be observed also when doing multi-condition training of the PLDA, but we can also see that it harmed the clean condition and helped more on the noisy one, which is an expected behavior.

Generative i-vector extraction training is unsupervised.  When we add augmented data to the training list, i-vector extraction is forced to reserve a portion of parameters for representation of variability of noise, reverberation and so it limits parameters for speaker variability.  In our supervised discriminative approach, we are forcing i-vector extractor to do the opposite.  The extractor is forced to distinguish the speakers, so it should decrease the unwanted variability and keep as many parameters of the $\tmatrix$-matrix to the speaker variability. It can also help to limit usage GMM components which are not useful for speaker separation.

At this point it is appropriate to discuss our results when compared with the current x-vector recipes.  We are fully aware that we do not reach the performance of x-vectors.  Results presented here can be directly compared to our previous work~\cite{xvec:ondran} focused on analyzing the performance of the state-of-the-art i-vector and x-vector systems on the very same datasets. Here we present the i-vector system that is based purely on MFCCs, while in~\cite{xvec:ondran} we were using concatenation of MFCCs and DNN bottleneck features. Our current plan is to discriminatively retrain the baseline system from~\cite{xvec:ondran} and then finally replace the i-vector component in the fully end-to-end system presented in~\cite{rohdin:icassp:2018} by the discriminatively trained i-vector extractor.

%\textcolor{red}{@@ Poznamka o tom, ze jde o initial experiment, ci proof of concept a ze jsme si toho vedomi, ze jde ziskat treba i lepsi baseline s BN-MFCC}

%%%%%%%%%%%%%%%%%%%%%%%%%%%%%%%%%%%%%%%%%%%%%%%%%%%%%%%%%%%%%%%%%%%%%%%%%%%%%%%%
%%%%%%%%%%%%%%%%%%%%%%%%%%%%%%%%%%%%%%%%%%%%%%%%%%%%%%%%%%%%%%%%%%%%%%%%%%%%%%%%
\subsection{Experiment observations}

%During our experiments, we had to deal with several issues: depth of the classifier, overtraining, setting learning rates, and regularization. 

We found out, that robust classifier was necessary for proper $\tmatrix$-matrix
retraining. 
We have conducted experiments with different depth of NN multi-class classifier until we settled on a topology with no hidden layer, which effectively equals to logistic regression.  With this setup, we avoid problems with overtraining (especially in the early stage of our endeavor, where we did not use augmented data), there are fewer parameters to train, and time and memory requirements are within reasonable limits, yielding an overall robust classifier.
%. The next benefit was fewer parameters to train, then also training time and memory consumption.  

%For effective i-vector extractor re-training, the well-trained classifier was necessary.  Poorly trained classifier (during the initial stage) went to the training, where i-vector extractor staid without change and the only classifier was training (during the training stage, where classifier and i-vector extractor were trained together).  This could be noticed also from the mean square distance between initial T matrix and actual T matrix state during training, where we observe negligible changes only. The similar observation was spotted with a more complicated classifier, where during the full pipeline training was focused mostly on the classifier. 

For effective i-vector extractor re-training, a well-trained classifier was crucial.  In stage-2 of the training (where classifier was jointly retrained with the $\tmatrix$-matrix), a poorly trained classifier resulted in either negligible or even harmful update of the $\tmatrix$-matrix. 

Because of its size, matrix $\mathbf{T}$ was prone to overtraining, therefore, 
regularization was necessary.  We have chosen L2 regularization centered around
the initial ML matrix $\mathbf{T}_{init}$.
% , resulting in the following objective function: 
% %
% %As the regularization, we use mentioned distance between the initial $\mathbf{T})$ matrix ($\mathbf{T})$ matrix trained with the generative algorithm) and our discriminatively retrained matrix. Our loss function was changed to:
% %
% \begin{eqnarray} 
% \label{Eq:reg}
% %E_{\mathrm{L2}}(\mathbf{W}, \mathbf{T}) = E(\mathbf{W}, \mathbf{T}) + \alpha E_{MSE}(\mathbf{T},\mathbf{T}_{init}),
% E_{\mathrm{L2}}(\mathbf{W}, \mathbf{T}) = E(\mathbf{W}, \mathbf{T}) + \alpha \left\|\mathrm{vec}(\tmatrix)-\mathrm{vec}(\tmatrix_{init})\right\|^{2},
% \end{eqnarray}
% %
% where $E(\mathbf{W}, \mathbf{T})$ is our multi-class cross entropy as defined in 
% Eq.~(\ref{Eq:cross}), $\alpha$ is the regularization coefficient, and 
% $\tmatrix_{init}$ is the initial ML matrix. 
This limits the estimate of $\mathbf{T}$ from moving too far from the initial 
(already well-estimated) state. 
%This way, optimizing the objective function fine tunes the already good parameters.

% Until this stage, we had the learning rate set to 0.1 (both for training of the classifier as well as for the full pipeline training). 
After several unsuccessful experiments, where the change of $\tmatrix$ was too rapid, we set learning rate during the full pipeline training to $10^{-3}$ (so far $10^{-1}$ was used). After this change, the regularization was not necessary anymore, and we received stable training.

Fixing the parameters of the classifier during stage-2 (and retraining only 
$\tmatrix$) led to minor effect on the system, compared to the joint training. 

%The last set of experiments, what we did was with the different type of optimizer. Next, to the stochastic gradient descent, we also used the Adam optimizer. Adam optimizer helps us train model faster compared to stochastic gradient descent.  These experiments were done only for 400-dim variant of the system, therefore we did not include them into the table. 

Retraining $\tmatrix$ from randomly initialized matrix rather than from a
ML estimate did not lead to convergence.

%The last set of experiments, where $\tmatrix$ was randomly initialized, did not lead to the successful training of fully discriminatively trained i-vector extractor. 
%
% These all experiments have finally led us to the presented simple procedure.

\section{Conclusion}

In this work, we have presented a way of refining a standard generative 
i-vector extractor via discriminative training.  We were able to outperform the generative baseline and make use of additional data obtained by the means of augmentation to further improve the performance when using the discriminative training.  Our approach conveniently fits to the current efforts of building a fully end-to-end discriminative systems, and provides a way for a robust initialization of such a large and important part of the system.
Needless to say, we have not created a new state-of-the-art system, however, we have prepared a solid platform for our further research.
\bibliographystyle{ieeebib}
\bibliography{main}

\begin{thebibliography}{10}

\bibitem{lozano_odyssey_2016}
A.~Lozano-Diez, A.~Silnova, P.~Mat{\v{e}}jka, O.~Glembek, O.~Plchot,
  J.~Pe{\v{s}}{\'{a}}n, L.~Burget, and J.~Gonzalez-Rodriguez,
\newblock ``{A}nalysis and {O}ptimization of {B}ottleneck {F}eatures for
  {S}peaker {R}ecognition,''
\newblock in {\em Proceedings of Odyssey 2016}. 2016, vol. 2016, pp. 352--357,
  International Speech Communication Association.

\bibitem{Lei_icassp_2014}
Y.~Lei, N.~Scheffer, L.~Ferrer, and M.~McLaren,
\newblock ``A novel scheme for speaker recognition using a phonetically-aware
  deep neural network,''
\newblock in {\em 2014 IEEE International Conference on Acoustics, Speech and
  Signal Processing (ICASSP)}, May 2014, pp. 1695--1699.

\bibitem{Novoselov_interspeech_2015}
S.~Novoselov, T.~Pekhovsky, O.~Kudashev, V.~S. Mendelev, and A.~Prudnikov,
\newblock ``{N}on-linear {PLDA} for i-vector speaker verification,''
\newblock in {\em 2017 IEEE International Conference on Acoustics, Speech and
  Signal Processing (ICASSP)}, Sept 2015, pp. 214--218.

\bibitem{Bhattacharya_SLT16}
G.~Bhattacharya, J.~Alam, P.~Kenny, and V.~Gupta,
\newblock ``Modelling speaker and channel variability using deep neural
  networks for robust speaker verification,''
\newblock in {\em 2016 {IEEE} Spoken Language Technology Workshop, {SLT} 2016,
  San Diego, CA, USA, December 13-16}, 2016.

\bibitem{Ghahabi_icassp_2014}
O.~Ghahabi and J.~Hernando,
\newblock ``Deep belief networks for i-vector based speaker recognition,''
\newblock in {\em 2014 IEEE International Conference on Acoustics, Speech and
  Signal Processing (ICASSP)}, May 2014, pp. 1700--1704.

\bibitem{Variani_icassp_2014}
E.~Variani, X.~Lei, E.~McDermott, I.~L. Moreno, and J.~Gonzalez-Dominguez,
\newblock ``Deep neural networks for small footprint text-dependent speaker
  verification,''
\newblock in {\em 2014 IEEE International Conference on Acoustics, Speech and
  Signal Processing (ICASSP)}, May 2014, pp. 4052--4056.

\bibitem{heighold_icassp_2016}
G.~Heigold, I.~Moreno, S.~Bengio, and N.~Shazeer,
\newblock ``End-to-end text-dependent speaker verification,''
\newblock in {\em 2016 IEEE International Conference on Acoustics, Speech and
  Signal Processing (ICASSP)}, March 2016, pp. 5115--5119.

\bibitem{zhang_slt_2016}
S.~X. Zhang, Z.~Chen, Y.~Zhao, J.~Li, and Y.~Gong,
\newblock ``{E}nd-to-{E}nd attention based text-dependent speaker
  verification,''
\newblock in {\em 2016 IEEE Spoken Language Technology Workshop (SLT)}, Dec
  2016, pp. 171--178.

\bibitem{snyder_slt_2016}
D.~Snyder, P.~Ghahremani, D.~Povey, D.~Garcia-Romero, Y.~Carmiel, and
  S.~Khudanpur,
\newblock ``{D}eep neural network-based speaker embeddings for end-to-end
  speaker verification,''
\newblock in {\em 2016 IEEE Spoken Language Technology Workshop (SLT)}, Dec
  2016, pp. 165--170.

\bibitem{Bhattacharaya_interspeech_2017}
G.~Bhattacharya, J.~Alam, and P.~Kenny,
\newblock ``{D}eep {S}peaker {E}mbeddings for {S}hort-{D}uration {S}peaker
  {V}erification,''
\newblock in {\em Interspeech 2017}, 08 2017, pp. 1517--1521.

\bibitem{snyder_interspeech_2017}
D.~Snyder, D.~Garcia-Romero, D.~Povey, and S.~Khudanpur,
\newblock ``{D}eep {N}eural {N}etwork {E}mbeddings for {T}ext-{I}ndependent
  {S}peaker {V}erification,''
\newblock in {\em Interspeech 2017}, Aug 2017.

\bibitem{xvec:Snyder2018}
David Snyder, Daniel Garcia-Romero, Greg Sell, Daniel Povey, and Sanjeev
  Khudanpur,
\newblock ``{X}-vectors: {R}obust {DNN} {E}mbeddings for {S}peaker
  {R}ecognition,''
\newblock in {\em Proceedings of ICASSP}, 2018.

\bibitem{dix:glembek}
Ond{\v{r}}ej Glembek, Luk{\'{a}}{\v{s}} Burget, Niko Br{\"{u}}mmer,
  Old{\v{r}}ich Plchot, and Pavel Mat{\v{e}}jka,
\newblock ``{D}iscriminatively {T}rained i-vector {E}xtractor for {S}peaker
  {V}erification,''
\newblock in {\em Proceedings of Interspeech 2011}. 2011, number~8, pp.
  137--140, International Speech Communication Association.

\bibitem{xvec:ondran}
Ond{\v{r}}ej Novotn{\'{y}}, Old{\v{r}}ich Plchot, Pavel Mat{\v{e}}jka, Ladislav
  Mo{\v{s}}ner, and Ond{\v{r}}ej Glembek,
\newblock ``{O}n the use of {X}-vectors for {R}obust {S}peaker {R}ecognition,''
\newblock in {\em Proceedings of Odyssey 2018}. 2018, number~6, pp. 168--175,
  International Speech Communication Association.

\bibitem{rohdin:icassp:2018}
Johan Rohdin, Anna Silnova, Mireia Diez, Old{\v{r}}ich Plchot, Pavel
  Mat\v{e}jka, and Luk{\'{a}}{\v{s}} Burget,
\newblock ``{E}nd-to-end {DNN} based speaker recognition inspired by i-vector
  and {PLDA},''
\newblock in {\em Proceedings of ICASSP}. 2018, IEEE Signal Processing Society.

\bibitem{DehakN_TASLP:2010}
N.~Dehak, P.~Kenny, R.~Dehak, P.~Dumouchel, and P.~Ouellet,
\newblock ``Front-{E}nd {F}actor {A}nalysis {F}or {S}peaker {V}erification,''
\newblock {\em IEEE Transactions on Audio, Speech, and Language Processing},
  vol. 19, no. 4, pp. 788--798, May 2011.

\bibitem{ferrer:sre11}
L.~Ferrer, H.~Bratt, L.~Burget, H.~Cernocky, O.~Glembek, M.~Graciarena,
  A.~Lawson, Y.~Lei, P.~Matejka, O.~Plchot, and N.~Scheffer,
\newblock ``{P}romoting robustness for speaker modeling in the community: the
  {PRISM} evaluation set,''
\newblock in {\em Proceedings of {SRE11} analysis workshop}, Atlanta, Dec.
  2011.

\bibitem{NIST_SRE:WWW}
``{N}ational {I}nstitute of {S}tandards and {T}echnology,''
  http://www.nist.gov/speech/tests/spk/index.htm.

\bibitem{SITW_evaluation_plan}
Mitchell McLaren, Luciana Ferrer, Diego Castan, and Aaron Lawson,
\newblock ``{T}he {S}peakers in the {W}ild ({SITW}) {S}peaker {R}ecognition
  {D}atabase,''
\newblock in {\em Interspeech 2016}, 2016, pp. 818--822.

\bibitem{NIST:SRE2016}
``The {NIST} year 2016 {S}peaker {R}ecognition {E}valuation {P}lan,'' \url{
  https://www.nist.gov/sites/default/files/documents/2016/10/\\07/sre16\_eval\_plan\_v1.3.pdf},
  2016.

\bibitem{MARDY:WWW}
``{M}ultichannel {A}coustic {R}everberation {D}atabase at {Y}ork,''
  http://www.commsp.ee.ic.ac.uk/~sap/resources/mardy-multichannel-acoustic-reverberation-database-at-york-database/.

\bibitem{AIR:WWW}
``{A}achen {I}mpulse {R}esponse {D}atabase,''
  http://www.iks.rwth-aachen.de/en/research/tools-downloads/databases/aachen-impulse-response-database/.

\bibitem{C4DM:WWW}
``{C4DM} ({C}enter for {D}igital {M}usic) {RIR} database,''
  http://isophonics.net/content/room-impulse-response-data-set.

\bibitem{C4DM:WWW2}
R.~Stewart and M.~Sandler,
\newblock ``{D}atabase of omnidirectional and {B}-format room impulse
  responses,''
\newblock in {\em 2010 IEEE International Conference on Acoustics, Speech and
  Signal Processing}, March 2010, pp. 165--168.

\bibitem{OpenAir:WWW}
``{O}pen{A}ir {I}mpulse {R}esponse {D}atabase,''
  http://www.openairlib.net/auralizationdb.

\bibitem{RVB:WWW}
``{R}everb {C}hallenge~~~~~~~,''
  http://reverb2014.dereverberation.com/index.html.

\bibitem{RWCP:WWW}
``{RWCP} {S}ound {S}cene {D}atabase,'' http://www.openslr.org/13/.

\end{thebibliography}

\end{document}